\documentclass[3p]{elsarticle}
\usepackage{amsmath}
\usepackage{comment}
\usepackage{color}
\usepackage{braket}
\usepackage{amssymb}

\def\Tr{\mathrm{Tr}}
\def\sdet{\mathrm{sdet}}

\newcommand \be{\begin{eqnarray}}
\newcommand \ee{\end{eqnarray}}

\def\simge{\mathrel{
    \rlap{\raise 0.511ex \hbox{$>$}}{\lower 0.511ex \hbox{$\sim$}}}}
\def\simle{\mathrel{
    \rlap{\raise 0.511ex \hbox{$<$}}{\lower 0.511ex \hbox{$\sim$}}}}

\begin{document}

\title{Diffusion method in Random Matrix Theory}
\author[a2]{Jacek Grela\corref{cor1}}
\ead{jacekgrela@gmail.com}
\address[a2]{M. Smoluchowski Institute of Physics and Mark Kac Complex Systems Research Centre, Jagiellonian University,  PL--30348 Krak\'ow, Poland}

\begin{abstract}
We introduce a simple yet powerful calculational tool useful in calculating averages of ratios and products of characteristic polynomials. The method is based on Dyson Brownian motion and Grassmann integration formula for determinants. It is intended as an alternative to other RMT techniques applicable to general gaussian measures. Resulting formulas are exact for finite matrix size $N$ and form integral representations convenient for large $N$ asymptotics. Quantities obtained by the method can be interpreted as averages over matrix models with an external source. We provide several explicit and novel calculations showing a range of applications.
\end{abstract}

\date{\today}

\maketitle

\section{Introduction}

One of the strengths of Random Matrix Theory lies in the abundance of powerful calculational tools, with the method of orthogonal polynomials \cite{MEHTA}, supersymmetric techniques \cite{EFETOV1} and free probability \cite{NICA1} among many others. This paper attempts to enlarge this family with a convenient technique we call the diffusion method. It serves as a fast and powerful framework useful when dealing with powers and ratios of characteristic polynomials averaged over external source gaussian measures. It began as a byproduct of QCD considerations made several years ago \cite{BN1} and was thereafter successfully applied to hermitian, Wishart and chiral models \cite{BNW1, BNW2, BGNW1}. The method uses a Dyson-like picture of dynamical matrices to achieve this goal.

Studying characteristic polynomials in the RMT community is now a prolific topic with many branches, its root however can be traced back to a remarkable relation connecting them to orthogonal polynomials
\begin{align*}
	\left < \det (z - X) \right >_X = \pi_N(z).
\end{align*}
Such objects were considered in \cite{BREZINHIKAMI1} and in many areas of application such as zeroes of Riemann $\zeta$ function \cite{KEATINGSNAITH1}, eigenvalue statistics in quantum chaotic systems \cite{ANDREEVSIMONS1} and matrix models of QCD \cite{VERBAARSCHOT1}. Moreover, products and ratios of characteristic polynomials reveal rich mathematical structures in both hermitian \cite{STRAHOVFYODOROV1,GUHR2} and non-hermitian \cite{AKEMANNVERNIZZI1, BGNTW1, BGNTW2} models. 

At the core of the method lies a seminal work of Dyson \cite{DYSON1} who observed that a static matrix model can be interpreted as a dynamical system. He showed that the joint probability density function for $N$ eigenvalues behaves exactly like a statistical system of $N$ "particles" interacting via logarithmic potential. Such particle system undergoes a Dysonian Brownian motion defined by the Langevin equation of the form:
\begin{align*}
	d \lambda_i = \sum_{j\neq j}\frac{1}{\lambda_i - \lambda_j} dt.
\end{align*}
This interacting Brownian motion is induced by a gaussian diffusion applied independently to matrix entries. Even though it lacks the harmonic potential usually present to attain the stationary the $\tau \to \infty$ limit, we show how to include this canonical approach of Dyson in the present framework. Studying RMT from the Brownian motion's point of view attracted attention of physicists \cite{BEENHAKKER1, NEUBERGER1} and mathematicians \cite{KATORI1, SPOHN1} alike.

Plan of the paper is as follows. In Sec. 2 we discuss the framework of the method in full generality and comment on its characteristics. In Sec. 3 we comment on the relation to standard random matrix models and on how to include the harmonic potential. In Sec. 4 we calculate five examples with special attention given to non-hermitian models. We show how to arrive at the formula for the averaged ratio of characteristic polynomials wrt. $\beta=2$ complex hermitian models and derive a novel duality-type equation for averaged products of characteristic polynomials in the complex $\beta=2$ non-hermitian ensemble. Furthermore, we compute a new integral representation of the averaged characteristic polynomial in a correlated variation of complex $\beta=2$ non-hermitian ensemble, compute the same object for a multiplication of two non-hermitian matrices and study real/complex non-hermitian ensemble crossover. The examples provided in the last section are mostly novel results.
%%%%%%%%%%%%%%%%%%%%%%%%%%%%%%%%%%%%%%%%%%%%%%%%%%%%%%%%%%%%%%%%%%%%%%
%%%%%%%%%%%%%%%%%%%%%%%%%%%%%%%%%%%%%%%%%%%%%%%%%%%%%%%%%%%%%%%%%%%%%%
\section{Diffusion method}

We introduce entrywise diffusive dynamics to $M$ - a $N\times N$ matrix of interest with prescribed symmetries. Well-suited formalism for our purpose is the multidimensional heat equation:
\begin{align}
\label{diff}
	\partial_\tau P(M,\tau) = \frac{1}{N} \Delta_M P(M,\tau),
\end{align}
where $P(M,\tau)$ is the joint probability density function, $\Delta_M$ denotes a Laplace operator over indepedent degrees of freedom of $M$ and the constant $1/N$ is a convention. For example, if $M=M^\dagger$, the hermitian laplacian is flat and reads $\Delta_{M} = \frac{1}{2} \sum_i \partial^2_{x_{ii}}+ \frac{1}{2} \sum_{i<j} (\partial^2_{x_{ij}} + \partial^2_{y_{ij}})$. A class of objects to which the method applies are the ratios and products of characteristic polynomials denoted in general by $D(Z,M)$. We are ultimately interested in formulas for the average
\begin{align}
	\overline{D_\tau}(Z) := \left < D(Z,M_\tau) \right >_{M_\tau} = \int dM P(M,\tau) D(Z,M) ,
\end{align}
where $\left < \right >_{M_\tau}$ is a normalized averaging over the dynamical matrices $M_\tau$. Second equality is a consequence of the defintion of jPDF $P(X,\tau) = \left < \delta (M_\tau-X) \right >_{M_\tau} $. 

To proceed, we extend $D(Z,M) \to \textbf{D}(Z,M;\Lambda)$ by introducing parameter-like variables $\Lambda$ such that $\lim_{\Lambda \to \Lambda_0} \textbf{D}(Z,M; \Lambda) = D(Z,M)$ with $\Lambda_0 = 0$ in most cases. This procedure in practice means we rewrite $D$ as a block determinant and introduce off-diagonal terms accordingly. The sole purpose of this auxillary construction is the search of diffusive equations in $\Lambda$-parameter space for the averaged deformed quantity $\overline{\textbf{D}_\tau}(Z;\Lambda) = \left < \textbf{D}(Z,M_\tau;\Lambda) \right >_{M_\tau}$.

Constructing useful deformation parameters $\Lambda$ is not difficult however it varies between problems and is found in a case-by-case manner. Usefulness is determined solely by its ability to produce a closed diffusive type equation for $\overline{\textbf{D}}_\tau$. In simplest cases introduction of additional parameters is superfluous since already the arguments $Z$ of charactetistic polynomials suffice in deriving the diffusion equation. 

In order to find the final equation we consider a time derivative of $\overline{\textbf{D}_\tau}$:
\begin{align}
	\partial_\tau \overline{\textbf{D}_\tau} = \frac{1}{N} \int dM \Delta_M P(M,\tau) \textbf{D}(Z,M;\Lambda) = \frac{1}{N} \int dM P(M,\tau) \Delta_M \textbf{D}(Z,M;\Lambda),
\end{align}
where we used the equation \eqref{diff} and integrated by parts to move the differential operator to $\textbf{D}$. Remaining task is to find $\boldsymbol{\Delta}_\Lambda$ such that the condition 
\begin{align}
\label{condition}
	\Delta_M \textbf{D}(Z,M;\Lambda) = \boldsymbol{\Delta}_\Lambda \textbf{D}(Z,M;\Lambda)
\end{align}
is satisfied. We then write the diffusive equation as
\begin{align}
\label{paramdiff}
	\partial_\tau \overline{\textbf{D}_\tau}(Z;\Lambda) = \frac{1}{N} \boldsymbol{\Delta}_\Lambda \overline{\textbf{D}_\tau}(Z;\Lambda).
\end{align}
As can be seen from the condition \eqref{condition}, gaussian laplacians on $M$ manifold transform into Gaussian laplacian operators on $\Lambda$ space but, at the same time, we observe a drastic decrease in the number of variables. 

Practical calculations solving \eqref{condition} use the Grassmann/complex representation of determinants or fermionization technique
\begin{align}
	\det M \sim \int d\  e^{-\Tr \eta^\dagger M \eta}, \qquad \det M^{-1} \sim \int d \alpha e^{-\Tr \alpha^\dagger M \alpha },
\end{align}
where the constants are not important for our purpose. In this representation the derivatives are exceptionally simple to compute. This is also the reason why it is desirable of $\Lambda$ deformations to retain determinantal structure. 

As a last step, we solve the equation \eqref{paramdiff} and set auxillary parameters back to their initial values $\Lambda \to \Lambda_0$ to recover the object of interest $\overline{D_\tau}$.
%%%%%%%%%%%%%%%%%%%%%%%%%%%%%%%%%%%%%%%%%%%%%%%%%%%%%%%%%%%%%%%%%%%%%%
%%%%%%%%%%%%%%%%%%%%%%%%%%%%%%%%%%%%%%%%%%%%%%%%%%%%%%%%%%%%%%%%%%%%%%
\subsection{Comments and resume}
The method is applicable to any type of entrywise diffusion \eqref{diff} which is perfect for models with nontrivial variance structure. An instance of such model was considered in Example 3. The diffusion equation in parameter space \eqref{paramdiff} has in general low dimensionality when compared to the matrix size, this duality phenomenon is widely observed in other methods. The resulting formulas, being the solution to a differential equation, are always dependent on the initial condition - the starting matrix $M_0$. For standard matrix models this is translated into a jPDF with an external source \cite{BREZINHIKAMI2}, an observation elucidated in Sec. 3. Final formulas can be viewed also as integral representations convenient for large $N$ analysis.

A general way to proceed follows these subsequent steps:
\begin{enumerate}
	\item introduce entrywise diffusion of choice \eqref{diff},
	\item define object $D$ (i.e. product and ratios of determinants) and form a $\Lambda$ parameter extension $\textbf{D}$,
	\item infer a diffusion equation in the $\Lambda$ space for the averaged quantity $\overline{\textbf{D}}$ with the help of Grassmann/complex representation of $\textbf{D}$ and condition \eqref{condition},
	\item solve the equation \eqref{paramdiff} and set $\Lambda$ parameters to its undeformed values $\Lambda_0$ to recover the object of interest.
\end{enumerate}
%%%%%%%%%%%%%%%%%%%%%%%%%%%%%%%%%%%%%%%%%%%%%%%%%%%%%%%%%%%%%%%%%%%%%%
%%%%%%%%%%%%%%%%%%%%%%%%%%%%%%%%%%%%%%%%%%%%%%%%%%%%%%%%%%%%%%%%%%%%%%
\section{Relation to standard random matrix models}
In previous section we have discussed a general framework in the diffusive language, here we comment on how to connect this approach to standard random matrix measures. Entrywise diffusion \eqref{diff} is, as a differential equation, implicitly reinforced with initial conditions which we consider to be of delta function type $P(M,\tau \to 0) = \delta(M-M_0)$. The matrix at time $\tau$ is equal to
\begin{align}
	M_\tau = M_0 + \sqrt{\tau} \mathcal{M},
\end{align}
where $\mathcal{M}$ is a matrix chosen randomly from jPDF $P_{\mathcal{M}_0=0}(\mathcal{M},\tau=1)$. By inverting this equation we conclude that
\begin{align}
	P(\mathcal{M}) d \mathcal{M} \sim P \left (\frac{M(\tau) - M_0}{\sqrt{\tau}} \right ) d M(\tau),
\end{align}
where on the RHS the time $\tau$ serves as a parameter and can be set to $\tau=1$. A family of matrix models of the form $P(M-M_0)$ are called external source or shifted mean models \cite{BREZINHIKAMI2}. In present context, averaging over dynamical matrices with time $\tau$ is equivalent to matrix models of variance proportional to $\tau$ and with an external source:
\begin{align}
	\int dM P(M,\tau) X(M) \sim \int dM P \left (\frac{M-M_0}{\sqrt{\tau}},\tau=1 \right ) X(M).
\end{align}
The proportionality constant depends on the form of $P$. In particular, if $M=M^\dagger=H$ is complex hermitian, the solution to \eqref{diff} reads
\begin{align}
\label{pherm}
	P(H,\tau) = C \exp \left ( -\frac{N}{2\tau} \Tr (H - H_0)^2 \right ),
\end{align}
with $C = (\frac{N}{2\pi \tau})^{\frac{N^2}{2}}$. In the case of $M=X$ complex without any additional symmetries we calculate that
\begin{align}
\label{pgin}
	P(X,\tau) = C' \exp \left (- \frac{N}{\tau} \Tr (X-X_0)^\dagger (X-X_0) \right ),
\end{align}
with $C' = \left ( \frac{N}{\pi \tau} \right )^{N^2}$ so that $\int P dX = 1$. We observe how the time $\tau$ serves now as scale of the variance and the external source being the initial matrix $X_0 (H_0)$. 

We now address the lack of harmonic potential in the entrywise diffusion \eqref{diff} present in most treatments of Dyson Brownian motion. It serves as an effective confining well giving a non-zero stationary $\tau \to \infty$ limit. Its presence can be however mimicked by a suitable reparametrization \cite{BBCL1}.  Namely, by setting the source  $ M_0 \to M_0 e^{-\tau}$ and time variable $\tau \to \frac{1 - e^{-2\tau}}{2}$ we effectively retrieve a diffusion in the harmonic potential. As $\tau \to \infty$, the initial matrix information is lost as the diffusion fills the whole confining potential which was already observed in Dyson's paper \cite{DYSON1}.

%%%%%%%%%%%%%%%%%%%%%%%%%%%%%%%%%%%%%%%%%%%%%%%%%%%%%%%%%%%%%%%%%%%%%%
%%%%%%%%%%%%%%%%%%%%%%%%%%%%%%%%%%%%%%%%%%%%%%%%%%%%%%%%%%%%%%%%%%%%%%
\section{Examples}
This section is devoted to several interesting examples. It serves also as a tour-de-force showing the framework at work to calculate some new results and compare to known ones. The majority of them deal with the non-hermitian ensembles. 

Example 1 is devoted to the probably most thoroughly studied random matrix ensemble of complex $\beta=2$ hermitian matrices. We show the applicability of our method to the averaged ratio of determinants, obtain an integral representation for any external source $H_0$ and show how it reduces to known results \cite{STRAHOVFYODOROV1} for $H_0 \to 0$.
 
Example 2 elucidates on a certain duality-type formula for complex $\beta=2$ non-hermitian ensembles, a result which continues the successful programme of dualities obtained in hermitian \cite{FORRESTERWITTE1, DESROSIERS1} and non-hermitian \cite{AKEMANNVERNIZZI1} ensembles.

Example 3 is a calculation in the deformed Ginibre of type considered in \cite{FORRESTERRAINS1}, a model inspired by the doubly-degenerate Wishart ensemble \cite{WALTNER1, MCKKAY1}. We compute an integral representation and compare to known results in the vanishing external source limit.

Example 4 serves as a proof-of-concept in applying the method to multiplication of matrices which attracted a lot of attention recently \cite{BURDAAKEMANN1}. In this case we calculate an integral representation for the averaged characteristic polynomial. 

Last example is a toy-model used to study the crossover between real $\beta=1$ and complex $\beta=2$ Ginibre ensemble inspiried by the elliptic ensemble modelling the GUE-Ginibre transition. We arrive at the large $N$ formula of the real-axis bump developed as we vary the crossover parameter.

%%%%%%%%%%%%%%%%%%%%%%%%%%%%%%%%%%%%%%%%%%%%%%%%%%%%%%%%%%%%%%%%%%%%%%
%%%%%%%%%%%%%%%%%%%%%%%%%%%%%%%%%%%%%%%%%%%%%%%%%%%%%%%%%%%%%%%%%%%%%%
\subsection{Ratio of determinants for complex $\beta=2$ hermitian matrices}
In this example we show how to obtain formulas for averaged ratio of determinants by diffusion meethod. First, we introduce diffusion of the form
\begin{align}
	\partial_\tau P(H,\tau) = \frac{1}{2N} \left (\sum_k \partial_{x_{kk}}^2 + \sum_{i<j} \left ( \partial_{x_{ij}}^2 + \partial_{y_{ij}}^2\right ) \right ) P(H,\tau),
\end{align}
where by convention we have $H_{ij} = \frac{1}{\sqrt{2}} (x_{ij} + i y_{ij})$ for $i\neq j$ and $H_{ii} = x_{ii}$.
We consider the ratio of characteristic polynomials:
\begin{align}
	D(z,w,H)=\frac{\det (z-H)}{\det (w-H)}.
\end{align}
The extension to $\textbf{D}$ consists of two Grassmann parameters $p,q$ forming a super-determinant:
\begin{align}
\textbf{D}(z,w,H;q,p) = \sdet \left ( \begin{matrix} w-H & q \\ p & z-H  \end{matrix} \right ) = \frac{\det (z-H - p(w-H)^{-1} q)}{\det (w-H)}.
\end{align}
To solve condition \eqref{condition} we open the superdeterminant with anti- ($\eta$) and commuting ($\xi$) part:
\begin{align*}
\textbf{D} & \sim \int d\xi d\eta \exp \left [ - \left ( \begin{matrix} \bar{\xi} & \bar{\eta} \end{matrix} \right ) \left ( \begin{matrix} w-H & q \\ p & z-H  \end{matrix} \right ) \left ( \begin{matrix} \xi \\ \eta \end{matrix} \right ) \right ] = \int d\xi d\eta e^{T_G}, \\
	T_G & = \sum_i \left ( x_{ii} (\bar{\eta}_i \eta_i + \bar{\xi}_i \xi_i ) -z \bar{\eta}_i \eta_i - w \bar{\xi}_i \xi_i - \bar{\eta}_i p \xi_i - \bar{\xi}_i q \eta_i \right ) + \\
	& + \frac{1}{\sqrt{2}} \sum_{i<j}  x_{ij} (\bar{\eta}_i \eta_j - \eta_i \bar{\eta}_j+\bar{\xi}_i \xi_j + \xi_i \bar{\xi}_j ) + \frac{i}{\sqrt{2}} \sum_{i<j} y_{ij} (\bar{\eta}_i \eta_j + \eta_i \bar{\eta}_j + \bar{\xi}_i \xi_j - \xi_i \bar{\xi}_j),
\end{align*}
where the exact proportionality factor is not needed in the calculation so we omit it.
We calculate the LHS of \eqref{condition} as
\begin{align}
	\Delta_H \textbf{D} = \frac{1}{N} \int d\xi d\eta \left ( \sum_i \bar{\eta}_i \eta_i \bar{\xi}_i \xi_i + \frac{1}{2} \sum_i \bar{\xi}_i^2 \xi_i^2 + \sum_{i<j} (\xi_i \bar{\xi}_j - \eta_i \bar{\eta}_j)(\bar{\eta}_i \eta_j + \bar{\xi}_i \xi_j) \right ) e^{T_G},
\end{align}
and we infer the laplacian $\boldsymbol{\Delta}_\Lambda$ as
\begin{align}
	\boldsymbol{\Delta}_{\Lambda} = \frac{1}{2N} \left ( \partial_{ww} - \partial_{zz} - 2 \partial_p \partial_q \right ).
\end{align}
The final diffusion equation is therefore equal
\begin{align}
\label{ex1sol}
 \partial_\tau \overline{\textbf{D}_\tau}(z,w;p,q) = \frac{1}{2N} \left ( \partial_{ww} - \partial_{zz} - 2\partial_p \partial_q \right ) \overline{\textbf{D}_\tau}(z,w;p,q) .
\end{align}
We comment on two peculiarities of \eqref{ex1sol}. In the $z$ direction it has a negative diffusivity constant and the diffusion happens also in the $p,q$ Grassmann "directions". Although the negative diffusion constant is not very physical, in the RMT context it is known to be responsible for universal oscillatory type behaviour \cite{BGNW1}. To deal with it on a technical level we simply Wick-rotate the $z$ variable. As to the Grassmann diffusion we make use of the "flatness" property $q^2=0, p^2=0$ and expand $\textbf{D}(z,w,H;p,q) = \textbf{D}^{(1)} + p \textbf{D}^{(2)} + q \textbf{D}^{(3)} + qp \textbf{D}^{(4)}$. Using this decomposition we rewrite \eqref{ex1sol} as an equivalent system of 4 equations for each $\textbf{D}^{(i)}$:
\begin{align}
	\partial_\tau \overline{\textbf{D}_\tau}^{(1)} & = \frac{1}{2N} \left ( \partial_{ww} - \partial_{zz} \right ) \overline{\textbf{D}_\tau}^{(1)} - \frac{1}{N} \overline{\textbf{D}_\tau}^{(4)} \label{ex1sol1}, \\
	\partial_\tau \overline{\textbf{D}_\tau}^{(2)} & = \frac{1}{2N} \left ( \partial_{ww} - \partial_{zz} \right )\overline{\textbf{D}_\tau}^{(2)} \label{ex1sol2}, \\
	\partial_\tau \overline{\textbf{D}_\tau}^{(3)} & = \frac{1}{2N} \left ( \partial_{ww} - \partial_{zz} \right ) \overline{\textbf{D}_\tau}^{(3)} \label{ex1sol3}, \\
	\partial_\tau \overline{\textbf{D}_\tau}^{(4)} & = \frac{1}{2N} \left ( \partial_{ww} - \partial_{zz} \right ) \overline{\textbf{D}_\tau}^{(4)} \label{ex1sol4}.
\end{align}
To find the solution of \eqref{ex1sol} we first observe that only equations \eqref{ex1sol1} and \eqref{ex1sol4} contain relevant components $i=1,4$. This is evident by observing that ultimately we are interested in the limit $\lim_{p,q\to 0}\textbf{D} = \textbf{D}^{(1)}$. We use a kernel of the heat equation
\begin{align}
	& K_\tau(z,w;y,v) = \frac{N}{2\pi \tau} \exp \left ( - \frac{N}{2\tau} (v-w)^2 - \frac{N}{2\tau} (y-iz)^2 \right ).
\end{align}
The solution to \eqref{ex1sol4} is
\begin{align}
	\overline{\textbf{D}_\tau}^{(4)}(z,w) = \int dy dv K_\tau(z,w;y,v) \textbf{D}_0^{(4)}(-iy,v;H_0) =: \left ( K_\tau \circ \textbf{D}_0^{(4)} \right )(z,w),
\end{align}
with $H_0$ denoting the initial matrix and introducing the convolution operator $\circ$. With this notation, the solution to the inhomogeneous heat equation \eqref{ex1sol1} is
\begin{align}
\label{ex1soll}
	\overline{\textbf{D}_\tau}^{(1)}(z,w) = \left (K_\tau \circ \left ( \textbf{D}^{(1)}_0 - \frac{\tau}{N} \textbf{D}_0^{(4)} \right ) \right )(z,w) = \left (K_\tau \circ \textbf{D}^{(1)}_0 \right )(z,w) - \frac{\tau}{N} \overline{\textbf{D}_\tau}^{(4)}(z,w) .
\end{align}
To write explicitly the solution we expand the initial condition
\begin{align*}
	& \textbf{D} = \frac{\det(z-H_0)}{\det(w-H_0)} \left ( 1 + qp \Tr \frac{1}{(z-H_0)(w-H_0)} \right ), \\
	& \textbf{D}_0^{(1)}(y,v;H_0) = \frac{\det(y-H_0)}{\det(v-H_0)}, \qquad \textbf{D}_0^{(4)}(y,v;H_0) = \frac{\det(y-H_0)}{\det(v-H_0)} \Tr \frac{1}{(y-H_0)(v-H_0)} .
\end{align*}
We set diagonal matrix $H_0 = \textrm{diag}(h_1,...,h_N)$, $\vec{h}= (h1,...,h_N)$ and introduce
\begin{align}
	\pi_{\vec{h}}(z) = \sqrt{\frac{N}{2\pi \tau}} \int du e^{-\frac{N}{2\tau} (u-iz)^2} \prod_{i=1}^N(-iu-h_i), \\
	\theta_{\vec{h}}(w) = \sqrt{\frac{N}{2\pi \tau}} \int dq e^{-\frac{N}{2\tau}(q-w)^2} \prod_{i=1}^N \frac{1}{(q-h_i)} .
\end{align}
The averaged ratio of characteristic polynomials \eqref{ex1soll} is equal to
\begin{align}
\label{ex1fin}
	\overline{D_\tau}(z,w) = \pi_{\vec{h}}(z) \theta_{\vec{h}}(w) - \frac{\tau}{N} \sum_{i=1}^N \pi_{\vec{h}-\vec{e}_i}(z) \theta_{\vec{h}+\vec{e}_i}(w),
\end{align}
with $\vec{e}_i = (0...0,1,0,..0)$ denoting a unit $N$-dimensional vector in the $i$-th direction. Using \eqref{pherm}, we write the average explicitly as
\begin{align}
	\overline{D_\tau}(z,w) = C \int dH e^{-\frac{N}{2\tau} \Tr (H-H_0)^2)} \frac{\det(z-H)}{\det(w-H)}.
\end{align}
These types of averages are present as building blocks of biorthogonal structures \cite{DESROSIERSFORRESTER1,BLEHERKUIJLAARS1} where $\pi$ and $\theta$ are the multiple orthogonal polynomials of type I and II respectively.

To recover known formulas for the GUE case, we set $h_i=0$. Then $\theta_k(w) = \gamma_{k-1} f_{k-1}(w)$ is proportional to the Cauchy transform $f_{k}(z) = \int \frac{e^{-\frac{N}{2\tau}s^2}}{z-s} \pi_k(s)$ and $\gamma_k = \frac{1}{k!} \left ( \frac{N}{\tau} \right )^k \sqrt{\frac{N}{2\pi \tau}}$. Along with  $\gamma_N/\gamma_{N-1} = \frac{1}{\tau}$, we rewrite \eqref{ex1fin} as:
\begin{align}
	\overline{D_\tau}(z,w) = \gamma_{N-1} \pi_N(z) f_{N-1}(w) - \tau \gamma_N f_N(w) \pi_{N-1}(z),
\end{align}
which is the ratio formula calculated for GUE ensemble in \cite{STRAHOVFYODOROV1}. 
%%%%%%%%%%%%%%%%%%%%%%%%%%%%%%%%%%%%%%%%%%%%%%%%%%%%%%%%%%%%%%%%%%%%%%
%%%%%%%%%%%%%%%%%%%%%%%%%%%%%%%%%%%%%%%%%%%%%%%%%%%%%%%%%%%%%%%%%%%%%%
\subsection{Duality formula for complex $\beta=2$ non-hermitian matrices}
Let $X_{ij} = x_{ij} + i y_{ij}$ be a complex $N\times N$ non-hermitian matrix. We introduce diffusive dynamics of the form
\begin{align}
	\partial_\tau P(X,\tau) = \frac{1}{N} \Tr \partial_{XX^\dagger} P(X,\tau) = \frac{1}{4N} \sum_{i,j} \left ( \partial^2_{x_{ij}} + \partial^2_{y_{ij}} \right ) P (X,\tau).
\end{align}
We aim at calculating an averaged product of characteristic polynomials
\begin{align}
	D^{(k)}(\mathcal{Z},X) = \det \left [\prod_{i=1}^k (z_i - X) (\bar{z}_i - X^\dagger) \right ].
\end{align}
for which a deformation is a $2kN$ block matrix of the form
\begin{align}
	\textbf{D}^{(k)}(\mathcal{Z},X;A) = \det \left ( \begin{matrix} \mathcal{Z}\otimes 1_N - 1_k \otimes X & -A^\dagger \otimes 1_N \\ A\otimes 1_N & \mathcal{Z}^\dagger \otimes 1_N - 1_k \otimes X^\dagger \end{matrix}  \right ),
\end{align}
where $\mathcal{Z} = \textrm{diag} (z_1, ... z_k)$ and $A$ is a complex $k\times k$ matrix representing the $\Lambda$-parameter space. We baptise $\textbf{D}^{(k)}$ the k-extended averaged characteristic polynomial (k-EACP) in agreement with \cite{BGNTW1} where authors considered particular case of $k=1$. In the limit $A \to 0$ we recover the $D^{(k)}$. 

To proceed further, we open the $\textbf{D}^{(k)}$ using Grassmann variables
\begin{align*}
 & \textbf{D}^{(k)}(\mathcal{Z},A,X) = \int d\eta d\xi ~e^{T_G}, \\
 T_G & = \sum_{i=1}^k \bar{\eta}^{(i)} \cdot \eta^{(i)} z_i + \sum_{i=1}^k \bar{\xi}^{(i)} \cdot \xi^{(i)} \bar{z}_i - \sum_{i=1}^k \bar{\eta}^{(i)} \cdot X \cdot \eta^{(i)} - \sum_{i=1}^k \bar{\xi}^{(i)} \cdot X^\dagger \cdot \xi^{(i)} - \sum_{i,j} \bar{\eta}^{(i)} \cdot \xi^{(j)} A^\dagger_{ij} + \sum_{i,j} \bar{\xi}^{(i)} \cdot \eta^{(j)} A_{ij}
\end{align*}
with $kN$ dimensional vectors $\xi^{(i)}_j$  and $\eta^{(i)}_j$ ($i=1...k,j=1...N$) and $"\cdot"$ denotes a sum over $N$ dimensional indices. We set also $A_{xy} = a_{xy} + i b_{xy}$. 
We find that
\begin{align}
&	\Delta_X \textbf{D}^{(k)} = \frac{1}{N} \int d\eta d\xi \left ( \sum_{i,j=1}^k  \eta^{(i)} \cdot \bar{\xi}^{(j)} \bar{\eta}^{(i)} \cdot \xi^{(j)} \right ) e^{T_G} ,
\end{align}
which determines the parameter laplacian as
\begin{align}
	\boldsymbol{\Delta}_\Lambda = \frac{1}{N} \Tr \partial_{A A^\dagger} = \frac{1}{4N}  \sum_{i,j} \left ( \partial^2_{a_{ij}} + \partial^2_{b_{ij}} \right ).
\end{align}
We thus arrive at the final equation for the average $\left < \textbf{D}^{(k)} \right >_\tau = \overline{\textbf{D}_\tau}^{(k)}(\mathcal{Z},A)$:
\begin{align}
	\partial_\tau \overline{\textbf{D}_\tau}^{(k)}(\mathcal{Z},A) = \frac{1}{N} \Tr \partial_{A A^\dagger} \overline{\textbf{D}_\tau}^{(k)}(\mathcal{Z},A),
\end{align}
where we observe a reduction in dimensionality to $k\times k$ diffusion. The solution to this equation is written explicitly as
\begin{align}
	\overline{\textbf{D}_\tau}^{(k)}(\mathcal{Z},X_0;A) = \left ( \frac{N}{\pi \tau} \right )^{k^2} \int dB e^{-\frac{N}{\tau}\Tr (B - A)(B^\dagger - A^\dagger)} \textbf{D}^{(k)}(\mathcal{Z},X_0;B),
\end{align}
where $A$ and $X_0$ are the initial values of parameter- and the randomized matrix respectively. We turn to the product of characteristic polynomials by setting $A \to 0$:
\begin{align}
\label{rhs}
	\overline{D_\tau}^{(k)}(Z,X_0) = \left ( \frac{N}{\pi \tau} \right )^{k^2} \int dB e^{-\frac{N}{\tau}\Tr B B^\dagger} \textbf{D}^{(k)}(\mathcal{Z},X_0;B).
\end{align}
To arrive at the duality formula, we use the explicit jPDF formula \eqref{pgin} on the average $\overline{D_\tau}^{(k)}$:
\begin{align}
\label{lhs}
	\overline{D_\tau}^{(k)}(Z,X_0) = C' \int dX e^{-\frac{N}{\tau} \Tr (X-X_0)^\dagger (X-X_0)} D^{(k)}(\mathcal{Z},X).
\end{align} 
We can therefore write the duality from \eqref{rhs} and \eqref{lhs}:
\begin{align}
	C' \int dX e^{-\frac{N}{\tau} \Tr XX^\dagger} \textbf{D}^{(k)}(\mathcal{Z},X+X_0;A=0) = \left ( \frac{N}{\pi \tau} \right )^{k^2} \int dB e^{-\frac{N}{\tau}\Tr B B^\dagger} \textbf{D}^{(k)}(\mathcal{Z},X_0;B),
\end{align}
with the definition repeated for clarity:
\begin{align}
	\textbf{D}^{(k)}(\mathcal{Z},X;A) = \det \left ( \begin{matrix} \mathcal{Z}\otimes 1_N - 1_k \otimes X & -A^\dagger \otimes 1_N \\ A\otimes 1_N & \mathcal{Z}^\dagger \otimes 1_N - 1_k \otimes X^\dagger \end{matrix}  \right ).
\end{align}
This new result is an extension of a similar formula for $X_0=0$ obtained in \cite{AKEMANNVERNIZZI1}. Such duality type formulas were studied in hermitian ensembles by \cite{FORRESTERWITTE1}, for general $\beta$ in \cite{DESROSIERS1} and in the context of string theory by \cite{KIMURA1} among others. 
%%%%%%%%%%%%%%%%%%%%%%%%%%%%%%%%%%%%%%%%%%%%%%%%%%%%%%%%%%%%%%%%%%%%%%
%%%%%%%%%%%%%%%%%%%%%%%%%%%%%%%%%%%%%%%%%%%%%%%%%%%%%%%%%%%%%%%%%%%%%%
\subsection{Deformed complex $\beta=2$ non-hermitian matrix}
In another example we deal with a variance deformed version of non-hermitian matrix model. Consider a standard diffusion equation
\begin{align}
\label{diffex3}
	\partial_\tau P(X,\tau) = \frac{1}{N} \sum_{i,j} \partial_{X_{ij}} \partial_{X^\dagger_{ji}} P(X,\tau),
\end{align}
\begin{comment}
\begin{align}
\label{diffex3}
	\partial_\tau P(X,\tau;\Gamma,\Omega) = \frac{1}{N} \sum_{i,j} \Gamma_{ii} \partial_{X_{ij}} \Omega_{jj} \partial_{X^\dagger_{ji}} P(X,\tau;\Gamma,\Omega)
\end{align}
which is a generalization of the time and space correlated matrix model used widely in applications $P \sim \exp \left (- \Tr \Gamma^{-1} X \Omega^{-1} X^\dagger \right )$ where matrices $\Gamma, \Omega$ are positive definite. In the definition of dynamics \eqref{} we have utilized the unitary transformation to bring correlation matrices to diagonal form. One can check by explicit calculation that solution to \eqref{diffex3} reads
\begin{align}
	P(X,\tau;\Gamma,\Omega) = \frac{C(\Gamma,\Omega)}{\tau^{N^2}} \exp \left ( - \frac{N}{\tau} \Tr \Gamma^{-1} (X-X_0) \Omega^{-1} (X^\dagger-X_0^\dagger) \right )
\end{align} 
with normalization constant $C$. 
\end{comment}
We are however interested in a slightly modified determinant of the form
\begin{align}
\label{objex3}
	D^{(\Gamma,\Omega)}(z,X) = \det \left (z - \Gamma^{-1} X \Omega^{-1} \right ) \det \left (\bar{z} - \Omega^{-1} X^\dagger \Gamma^{-1} \right ).
\end{align}
This peculiar definition introduces a deformation of the usual non-hermitian model inspired by Wishart ensembles with time ($\Gamma$) and space ($\Omega$) correlations \cite{MCKKAY1}. It arises by changing $X \to \Gamma X \Omega $ in both \eqref{diffex3} and \eqref{objex3} 
\begin{align}
\label{diffcorr}
	\tilde{P}(X,\tau) \sim P(\Gamma X \Omega,\tau) = \exp \left ( - \frac{N}{\tau} \Tr \Gamma^{2} (X-\tilde{X}_0) \Omega^{2} (X^\dagger-\tilde{X}_0^\dagger) \right ),
\end{align}
where $\tilde{X}_0 = \Gamma^{-1} X_0 \Omega^{-1}$. With such reparametrization the characteristic polynomials \eqref{objex3} averaged over standard non-hermitian diffusion \eqref{objex3} becomes a standard characteristic polynomials averaged over doubly-correlated diffusion of type \eqref{diffcorr}. To retain probabilistic interpretation, correlation matrices $\Gamma$ and $\Omega$ are positive definite and can be set diagonal. The extended determinant $\textbf{D}^{(\Gamma,\Omega)}$ reads
\begin{align}
	\textbf{D}^{(\Gamma,\Omega)}(z,X;w) = \det \left ( \begin{matrix} z - \Gamma^{-1} X \Omega^{-1} & - \Gamma^{-2} \bar{w} \\ \Omega^{-2} w & \bar{z} -  \Omega^{-1} X^\dagger \Gamma^{-1} \end{matrix} \right ),
\end{align}
with $\lim_{w\to 0} \textbf{D}^{(\Gamma,\Omega)} = D^{(\Gamma,\Omega)}$. We open it as
\begin{align*}
	& \textbf{D}^{(\Gamma,\Omega)}(z,X;w) \sim \int d\eta d\xi ~e^{T_G} , \\
	& T_G = \bar{\eta} \cdot \eta z + \bar{\xi} \cdot \xi \bar{z} - \bar{\eta} \cdot (\Gamma^{-1} X \Omega^{-1}) \cdot \eta - \bar{\xi} \cdot \Omega^{-1}X^\dagger \Gamma^{-1} \cdot \xi - \bar{w} \bar{\eta} \cdot \Gamma^{-2} \cdot \xi + w \bar{\xi} \cdot \Omega^{-2} \cdot \eta ,
\end{align*}
with $N$ dimensional Grassmann vectors $\xi_j$ and $\eta_j$ ($j=1...N$). Let the laplacian act on the deformed determinant
\begin{align}
	\Delta_X \textbf{D} = \frac{1}{N} \int d \eta d \xi \left ( \sum_{i,j} \Gamma_{ii}^{-2} \xi_i \bar{\eta}_i \Omega_{jj}^{-2} \bar{\xi}_j \eta_j \right ) e^{T_G} ,
\end{align}
by condition \eqref{condition} we infer that
\begin{align}
\boldsymbol{\Delta}_\Lambda = \frac{1}{N} \partial_{w\bar{w}} .
\end{align}
The final equation for the averaged  $\left < \textbf{D} \right >_\tau = \overline{\textbf{D}_\tau}$ reads
\begin{align}
	\partial_\tau \overline{\textbf{D}_\tau}(z,A) = \frac{1}{N} \partial_{w\bar{w}}\overline{\textbf{D}_\tau}(z,A) .
\end{align}
The equation is two dimensional so the solution reads, after $w \to 0$
\begin{align}
\label{sol3}
	\overline{D_\tau}(z,X_0) & = \frac{N}{\pi \tau} \int d^2u e^{-\frac{N}{\tau} |u|^2} \textbf{D}^{(\Gamma,\Omega)}(z,X_0;u) , \\
	\textbf{D}^{(\Gamma,\Omega)}(z,X;w) & = \det \left ( \begin{matrix} z - \Gamma^{-1} X \Omega^{-1} & - \Gamma^{-2} \bar{w} \\ \Omega^{-2} w & \bar{z} -  \Omega^{-1} X^\dagger \Gamma^{-1} \end{matrix} \right ).
\end{align}
which is the sought integral representation valid for general $X_0$ and correlations $\Gamma,\Omega$. This studied average is given, using \eqref{pgin}, explicitly as
\begin{align}
	\overline{D_\tau}(z,X_0) = C' \int dX \exp \left ( - \frac{N}{\tau} \Tr (X-X_0)^\dagger (X-X_0) \right ) D^{(\Gamma,\Omega)}(z,X), 
\end{align}
In the special $X_0 \to 0$ limit, the solution \eqref{sol3} reproduces the result of \cite{FORRESTERRAINS1}:
\begin{align}
	\overline{D_\tau}(z) =\frac{2N}{\tau} \int_0^\infty d\rho \rho e^{-\frac{N}{\tau} \rho^2} \prod_{i=1}^N (|z|^2 + \rho^2 \Gamma_{ii}^{-2} \Omega_{ii}^{-2}).
\end{align} 
%%%%%%%%%%%%%%%%%%%%%%%%%%%%%%%%%%%%%%%%%%%%%%%%%%%%%%%%%%%%%%%%%%%%%%
%%%%%%%%%%%%%%%%%%%%%%%%%%%%%%%%%%%%%%%%%%%%%%%%%%%%%%%%%%%%%%%%%%%%%%
\subsection{Two complex Ginibre multiplied}
In this example we show that the method is suitable for product of matrices which recently had drawn much attention recently \cite{BURDAAKEMANN1}. We introduce two matrices undergoing an entrywise diffusion
\begin{align}
	\partial_\tau P = \frac{1}{N} \Tr \partial_{X_1 X_1^\dagger} P + \frac{1}{N} \Tr \partial_{X_2 X_2^\dagger} P,
\end{align}
and consider a determinant
\begin{align}
	D(z,X_1,X_2) = \det \left (z-X_1 X_2 \right ) \det \left (\bar{z} - (X_1 X_2)^\dagger \right ) .
\end{align}
To proceed with the method, the extension of $D$ is an object linearized in $X_i$'s:
\begin{align}
	\textbf{D}(z,X_1,X_2;u,v,w) = \det \left ( \begin{matrix} z & -\bar{w} & 0 & X_1 \\ v & \bar{z} & X_2^\dagger & 0 \\ 0 & X_1^\dagger & u & w \\ X_2 & 0 & -\bar{v} & \bar{u} \end{matrix} \right ),
\end{align}
\begin{comment}\begin{align}
	\textbf{D}(z,X_1,X_2;u,v,z) = \det \left ( \begin{matrix} Q & \mathcal{X}_- \\ \mathcal{X}_+ & R \end{matrix} \right )
\end{align}
where block matrices are
\begin{align}
	Q = \left ( \begin{matrix} z & - \bar{w} \\ w & \bar{z} \end{matrix} \right ), \qquad R = \left ( \begin{matrix} u & - \bar{v} \\ v & \bar{u} \end{matrix} \right ), \qquad \mathcal{X}_- = \left ( \begin{matrix} 0 & X_1 \\ X_2^\dagger & 0 \end{matrix} \right ), \qquad \mathcal{X}_+ = \left ( \begin{matrix} 0 & X_1^\dagger \\ X_2 & 0 \end{matrix} \right )
\end{align}
\end{comment}
where one can check that $\lim_{u\to 1, v,w \to 0} \textbf{D} = D$. We open the $4N$ deformed determinants with the help of $\xi_1, \xi_2, \eta_1$ and $\eta_2$: 
\begin{align*}
	& \textbf{D} = \int d \xi d\eta e^{T_G}, \\
	& T_G = \bar{\xi}_1 \cdot \xi_1 z + \bar{\xi}_2 \cdot \xi_2 \bar{z}  + \bar{\eta}_1 \cdot \eta_1 u + \bar{\eta}_2 \cdot \eta_2 \bar{u} - \bar{\xi}_1 \cdot \xi_2 \bar{w} + \bar{\xi}_2 \cdot \xi_1 v + \bar{\eta}_1 \cdot \eta_2 w - \bar{\eta}_2 \cdot \eta_1 \bar{v} + \\
	& + \bar{\xi}_1 \cdot X_1 \cdot \eta_2 + \bar{\xi}_2 \cdot X_2^\dagger \cdot \eta_1 + \bar{\eta}_1 \cdot X_1^\dagger \cdot \xi_2 + \bar{\eta}_2 \cdot X_2 \cdot \xi_1 .
\end{align*}
The laplacian acting on $\textbf{D}$ is
\begin{align}
	\Delta_{X_1,X_2} = \frac{1}{N} \int d \eta d\xi \left (\bar{\xi}_1 \cdot \xi_2 \eta_2 \cdot \bar{\eta}_1 + \bar{\eta}_2 \cdot \eta_1 \xi_1 \cdot \bar{\xi}_2 \right )e^{T_G} ,
\end{align}
which dictates the parameter laplace operator to be of the form
\begin{align}
	\boldsymbol{\Delta}_\Lambda = \frac{1}{N} \left ( \partial_{\bar{w},w} + \partial_{v,\bar{v}} \right ).
\end{align}
The final equation for averaged determinant reads
\begin{align}
	\partial_\tau \overline{\textbf{D}_\tau}(z;u,v,w) = \frac{1}{N} \left ( \partial_{\bar{w},w} + \partial_{\bar{v},v} \right ) \overline{\textbf{D}_\tau}(z;u,v,w),
\end{align}
We write down the solution for $v,w \to 0$ and $u \to 1$:
\begin{align}
	\overline{D_\tau} \left (z,(X_1)_0,(X_2)_0 \right ) = \left ( \frac{N}{\pi \tau} \right )^2 \int d^2w d^2 v e^{ - \frac{N}{\tau}(|w|^2 + |v|^2)} \textbf{D} \left (z,(X_1)_0, (X_2)_0;1,v,w \right ) ,
\end{align}
with the definition
\begin{align}
\label{ex4sol}
	\textbf{D}(z,X_1,X_2;u,v,w) = \det \left ( \begin{matrix} z & -\bar{w} & 0 & X_1 \\ v & \bar{z} & X_2^\dagger & 0 \\ 0 & X_1^\dagger & u & w \\ X_2 & 0 & -v & \bar{u} \end{matrix} \right ) .
\end{align}
As before, we investigate the vanishing source limit $(X_i)_0 \to 0$ where
\begin{align}
	\textbf{D}(z,0,0;u,v,w) = \left ( 1+ \bar{v} w \right )^N  \left ( |z|^2 + v\bar{w} \right )^N .
\end{align}
The angles of $w,v$ in \eqref{ex4sol} can be integrated out with the help of hypergeometric function
\begin{align}
	\overline{D_\tau}(z) = \left (\frac{2N}{\tau} \right )^2 \int_0^\infty dp dq ~qp e^{-\frac{N}{\tau} (q^2 + p^2)}(|z|^2 + q^2 p^2)^N {}_2 F_1 \left (\frac{1-N}{2},-\frac{N}{2},1,\frac{4|z|^2p^2q^2}{(|z|^2+p^2 q^2)^2} \right ),
\end{align}
and we simplified further by introducing $p^2 = t \alpha, q^2 = \frac{t}{\alpha}$ and integrating over $\alpha$'s:
\begin{align}
	\overline{D_\tau}(z) = \left (\frac{2N}{\tau} \right )^2 \int_0^\infty dt ~t K_0 \left (\frac{2Nt}{\tau} \right)(|z|^2 + t^2)^N {}_2 F_1 \left (\frac{1-N}{2},-\frac{N}{2},1,\frac{4|z|^2t^2}{(|z|^2+t^2)^2} \right ),
\end{align}
where the average, according to \eqref{pgin}, is explicitly given as
\begin{align}
	\overline{D_\tau}(z) = (C')^2 \int dX_1 dX_2 \exp \left ( -\frac{N}{\tau} \Tr ( X_1^\dagger X_1 + X_2^\dagger X_2 ) \right ) \det (z - X_1 X_2) \det (\bar{z} - (X_1 X_2)^\dagger ) .
\end{align}
To the best of our knowledge, this result was not considered previously.
%%%%%%%%%%%%%%%%%%%%%%%%%%%%%%%%%%%%%%%%%%%%%%%%%%%%%%%%%%%%%%%%%%%%%%
%%%%%%%%%%%%%%%%%%%%%%%%%%%%%%%%%%%%%%%%%%%%%%%%%%%%%%%%%%%%%%%%%%%%%%
\subsection{Real/complex non-hermitian crossover model}

The diffusive method is applicable also to $\beta=1$ real non-hermitian ensembles, we use it to show a simple crossover model between real and complex non-hermitian ensembles. An entrywise diffusion is defined by
\begin{align}
	\partial_\tau P(X,\tau) = \frac{1}{4N} \sum_{n,m} \left ( \frac{\partial^2}{\partial x_{nm}^2} + \alpha^2 \frac{\partial^2}{\partial y_{nm}^2} \right ) P(X,\tau) .
\end{align}
By varying $\alpha$ between $0 \leftrightarrow 1$ the transition between $\beta=1$ and $\beta=2$ happens. We are interested in characteristic polynomial in the standard form
\begin{align}
	D(z) = \det (z-X) \det (\bar{z} - X^\dagger),
\end{align}
with $N\times N$ matrix $X_{ij} = x_{ij} + i y_{ij}$. The extension is given by
\begin{align}
	\textbf{D}(z,X;w) = \left ( \begin{matrix} z - X & - \bar{w} \\ w & \bar{z} - X^\dagger \end{matrix} \right ) ,
\end{align}
which, using the same techniques as previously, we arrive at an equation for the average characteristic polynomial
\begin{align}
	\partial_\tau \overline{\textbf{D}_\tau}(z;w) = \frac{1+\alpha^2}{2N} \partial_{w\bar{w}} \overline{\textbf{D}_\tau}(z;w).
\end{align}
The solution, after taking the $w \to 0$ limit, reads
\begin{align}
\label{ex5sol}
	\overline{D_\tau}(z,X_0) = \frac{2N}{\tau} \int_0^\infty dr r e^{-\frac{2N}{\tau(1+\alpha^2)}r^2} \textbf{D}(z,r) .
\end{align}
which is valid for any external source $X_0$. For vanishing external source $X_0 \to 0$, the formula \eqref{ex5sol}  agrees with results for real and complex Ginibre ensemble \cite{AKEMANNPHILIPSSOMMERS1,BGNTW1}. 

We are interested in the crossover region as it develops a bump located near the real axis $z = i y N^{-1/4}$ as we also scale the crossover parameter $\alpha = a N^{-1/4} $:
\begin{align}
	\overline{D_\tau} \sim e^{-a^4/2} e^{-\frac{2 a^2 \eta^2}{\tau}} \textrm{erfc}\left (\frac{\sqrt{2}\eta^2}{\tau} - \frac{a^2}{\sqrt{2}} \right ),
\end{align}
centered around $\eta=0$. This deformation is interpreted as the precursor of a division between real and complex eigenvalues present in pure $\beta=1$ Ginibre ensemble. 
\section{Conclusions}

The presented method is shown to be universally applicable to general gaussian type random matrix models and serves as a calculational tool for obtaining ratios and products of characteristic polynomials. 

Examples considered present many new results of integral representations. In particular, a novel duality formula for products of characteristic polynomials in complex Ginibre matrices, a previously not considered characteristic polynomial for the product of Ginibre matrices and a real/complex Ginibre crossover model was considered. 

Main advantage of the method is a huge reduction in the degrees of freedom. It is applicable also to instances of statistical models with non-trivial covariance structures. Alongside, external source-type perturbations are present in this approach which broadens the general applicability of the method. 
\section{Acknowledgements}
Author acknowledges the support of the Grant DEC-2011/02/A/ST1/00119 of the National Centre of Science and the Australian Government Endeavour Fellowship during which this work was done. He would also thank P. J. Forrester, M. A. Nowak and P. Warchol for reading the manuscript and valuable comments.

\end{document}